%% file: main.tex
\documentclass[conference]{IEEEconf}

\input epsf
\usepackage{graphicx,subfigure}
\usepackage{cite}
\usepackage[numbers,sort&compress]{natbib}
\usepackage{stfloats}
\usepackage{multirow}
\usepackage{color}

\usepackage{url}       
\begin{document}

\title{\textbf{\Large Test Reuse Based on Adaptive Semantic Matching across Android Mobile Applications}}

\author{Shuqi Liu$^{1}$, Yu Zhou$^{1,*}$, Tingting Han$^{2}$, and Taolue Chen$^{2}$\\
	\normalsize $^{1}$College of Computer Science and Technology, Nanjing University of Aeronautics and Astronautics, Nanjing, China\\
	\normalsize $^{2}$Department of Computer Science and Data Science, Birkbeck, University of London, UK\\
	\normalsize \{liushuqi, zhouyu\}@nuaa.edu.cn, \{t.han, t.chen\}@bbk.ac.uk \\
	\normalsize *corresponding author

}

\maketitle
\begin{abstract}
Automatic test generation can help verify and develop the behavior of mobile applications. Test reuse based on semantic similarities between applications of the same category has been utilized to reduce the manual effort of Graphical User Interface (GUI) testing. However, most of the existing studies fail to solve the semantic problem of event matching, which leads to the failure of test reuse. To overcome this challenge, we propose TRASM ({\bfseries T}est {\bfseries R}euse based on {\bfseries A}daptive {\bfseries S}emantic {\bfseries M}atching), a test reuse approach based on adaptive strategies to find a better event matching across android mobile applications. TRASM first performs GUI events deduplication on the initial test set obtained from test generation, and then employs an adaptive strategy to find better event matching, which enables reusing the existing test. Preliminary experiments with comparison to baseline methods on 15 applications demonstrate that  TRASM can improve the precision of GUI event matching while reducing the failure of test reuse and the running time required for test reuse.

\end{abstract}
\IEEEoverridecommandlockouts

%
\IEEEpeerreviewmaketitle

\input{sections/Introduction}

\input{sections/Related}

\input{sections/TRASMD}
\input{sections/Evaluation}

\input{sections/Conclusion}



\section*{Acknowledgments}
The work is partially supported by the National Natural Science Foundation of China (No. 61972197), the Natural Science Foundation of Jiangsu Province (No. BK20201292), and the Collaborative Innovation Center of Novel Software Technology and Industrialization. 
T. Chen is partially supported by an oversea grant
from the State Key Laboratory of Novel Software Technology, Nanjing
University (KFKT2022A03), Birkbeck BEI School Project (EFFECT), National Natural Science Foundation of China
(NSFC) under Grants (No. 62072309, 62272397). 




%

\input{sections/References}
\end{document}

%% file: sections/Introduction.tex
\section{Introduction}\label{sec1}

Graphical User Interface (GUI) testing is commonly employed to verify and develop the behaviors of applications by designing and executing test cases of GUI applications~\cite{anand2012automated}. However, 
with the ever increasing functionalities in mobile applications, 
it takes more effort for developers to manually design GUI test cases (GUI test in short) \cite{joorabchi2013real, kochhar2015understanding, linares2017developers}, which in turn decreases the efficiency of testing processes.

\begin{figure}[htbp]
\centering

\subfigure[The existing test for To-Do List]
{
    \begin{minipage}[]{\linewidth}
        \centering
        \includegraphics[scale=0.25]{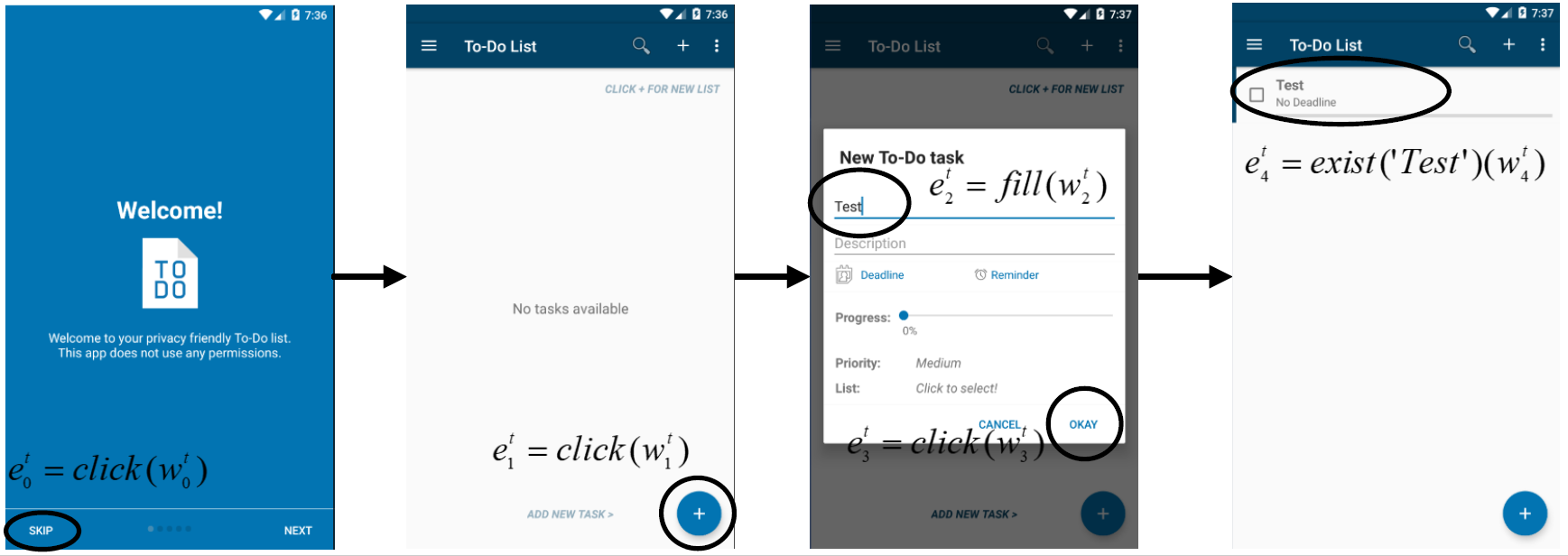}
    \end{minipage}
}
\subfigure[The reused test for Minimal]
{
 	\begin{minipage}[]{\linewidth}
        \centering
        \includegraphics[scale=0.25]{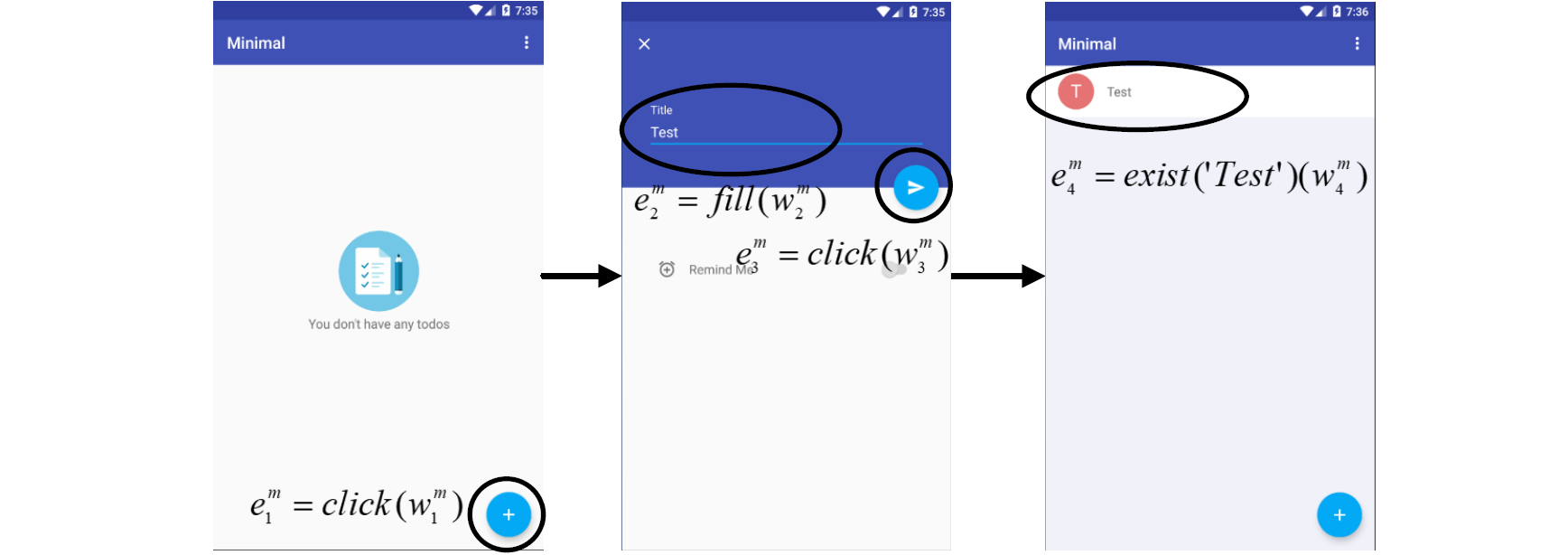}
    \end{minipage}
}
\caption{A simple example of test reuse. The test (b) is obtained by reusing the existing test (a). \label{fig1}}
\end{figure}

Considering the necessity of reducing time consumption, many researchers have conducted a series of investigation on automatic test generation \cite{gu2019practical, machiry2013dynodroid, mao2016sapienz,mirzaei2015sig,ermuth2016monkey,zhou2020user,dong2020time,memon2003gui,su2017guided,amalfitano2012using,wang2020combodroid} in GUI testing. Recently, some researchers observed that test reuse \cite{behrang2018automated,behrang2019test,behrang2020apptestmigrator,lin2019test,mao2021user,qin2019testmig,rau2018poster,rau2018transferring,mariani2021semantic} could be achieved by exploiting the semantic similarity of GUIs between similar applications to generate tests automatically. Figure~\ref{fig1} shows a simple example in which the existing test (a) of application {\itshape To-Do List} is successfully reused to application {\itshape Minimal}, and the reused test (b) is obtained. As Figure~\ref{fig1} shows, events  $e_{1}^{m}$,  $e_{2}^{m}$,  $e_{3}^{m}$, and  $e_{4}^{m}$ in test (b) are similar to  $e_{1}^{t}$,  $e_{2}^{t}$,  $e_{3}^{t}$, and  $e_{4}^{t}$ in test (a), respectively.

Existing research mainly focuses on how to accurately select specific characteristics of widgets in GUIs such as `text' and `resource-id'. Combining the selected characteristics, they design the semantic similarity calculation method between widgets to generate meaningful tests. They attempt to select widgets with high similarity in a similar application for matching each event of the existing test. However, little attention has been paid to optimizing the matching process. Taking Figure~\ref{fig1} as an example, the widget $w_{3}^{t}$ of {\itshape To-Do List} and the correctly similar widget $w_{3}^{m}$ of the application {\itshape Minimal} are laid out differently in the GUI. When reusing test (b) to the application {\itshape To-Do List}, adopting the existing approach may always incorrectly match the widget $w_{3}^{m}$ with other widgets. This may cause subsequent events to match incorrectly or even result in failed test reuse. In cases where the existing methods do not work well, it is necessary to adopt other corresponding measures. The lower the similarity of the generated event, the more likely the match is inappropriate. Hence, mining such events and exploring other widgets with more similar semantics to form events for substitution is considered.

In addition, in Figure~\ref{fig1}, the application {\itshape To-Do List} needs to skip the boot page that the application {\itshape Minimal} does not before entering the home page. And it is assumed that the event step is $e_{0}^{t}=click(w_{0}^{t})$. Under this assumption, the widget $w_{0}^{t}$ will match the widget with the highest similarity on the home page of the application {\itshape Minimal}. Obviously, the event produced by this step is redundant in the generated test. This simple example explains that we need to solve the event redundancy issue in the process of test reuse caused by some particular functionality in the existing test.

 Inspired by the above observation, in this paper, we propose a novel approach TRASM ({\bfseries T}est {\bfseries R}euse based on {\bfseries A}daptive {\bfseries S}emantic {\bfseries M}atching) to reuse the existing tests across android mobile applications. In addition, we carry out comparative experiments with the-state-of-the-art baseline approaches to evaluate our work. Overall, our main contributions are as follows:

\begin{enumerate}

\item 
We propose a novel approach TRASM, which utilizes an adaptive strategy to reuse more existing tests. TRASM can get more semantic matches in the generated test.

\item TRASM includes a GUI events deduplication method, which could eliminate duplicated events caused by reusing particular functionality contained in the existing test to improve the quality of the generated test.

\item We carry out extensive experiments which 
confirm that TRASM improves the accuracy of GUI event matching while reducing test reuse failures and reduces the running time required for test reuse.
\end{enumerate}

The rest of this paper is organized as follows. Section II introduces related work. Section III describes the main idea and the proposed approach in detail. Section IV carries out experimental evaluation. Finally, Section V concludes the paper and outlines future research.

%% file: sections/Related.tex
\section{Related work}
\subsection{Test Generation}
In order to improve the efficiency of developers, based on different exploration strategies, several studies on automatic test generation have been proposed, which has laid a solid foundation.

Sapienz \cite{mao2016sapienz} combined random fuzzing, systematic and search-based exploration, exploiting seeding and multi-level instrumentation to explore and optimize test sequences automatically. Gu et al. \cite{gu2019practical} dynamically abstracted the model by leveraging decision tree-based representation and updated the model by utilizing the evolution mechanism, which balances the accuracy and size of the model. ConmboDroid \cite{wang2020combodroid} obtained the use cases for verifying the unique functions of the application and then systematically enumerates and combines them to generate higher quality input. The advantage of their work is that they can mine more hidden bugs or achieve as high coverage as possible. Nevertheless, the test generated by their method is seldom standardized for verifying the application's functionality.

Different from their purpose and inspired by their exploration method, we focus on generating more meaningful tests based on semantic information.

 \subsection{Test Reuse}
Test reuse, as an alternative method to automatically generate GUI test, 
makes full use of existing resources to provide convenience for verifying the application's behavior.

Lin et al. proposed CRAFTDROID \cite{lin2019test}, an approach of test transfer across applications, which utilizes the GUI model extracted by static analysis to match event sequences similar to the semantics of the existing test in order. They realized the successful transfer of GUI and oracle events, which guides for improving test transfer. To more accurately express the similarity of widgets in test events, Mao et al. \cite{mao2021user} raised a semantic-based event fuzzy mapping strategy when matching candidate widgets to generate target events. They always greedily preferentially explore and match the widgets with the highest similarity. Unfortunately, when their similarity calculation method does not work well, the correctness of event matching will be threatened. Considering that the success of test reuse heavily depends on the semantic matching of test events, there is still space for improvement by adopting appropriate strategies to increase the quality of reused tests from the perspective of application functionality.

Leonardo et al. \cite{mariani2021semantic} conducted extensive research and pointed out that some attributes representing widgets play a negative role and how designing the semantic matching process is the most influential component to matched results. Their key findings point to an entry point for better reuse of test. Up to now, there is still no effective method to solve semantic problems \cite{choudhary2015automated, zeng2016automated}. Trying to optimize the generated test sequence to ensure the quality of reused tests should be an optional strategy.

%% file: sections/TRASMD.tex
\section{Our Approach}

\begin{figure*}[htbp]
\includegraphics[scale=0.4]{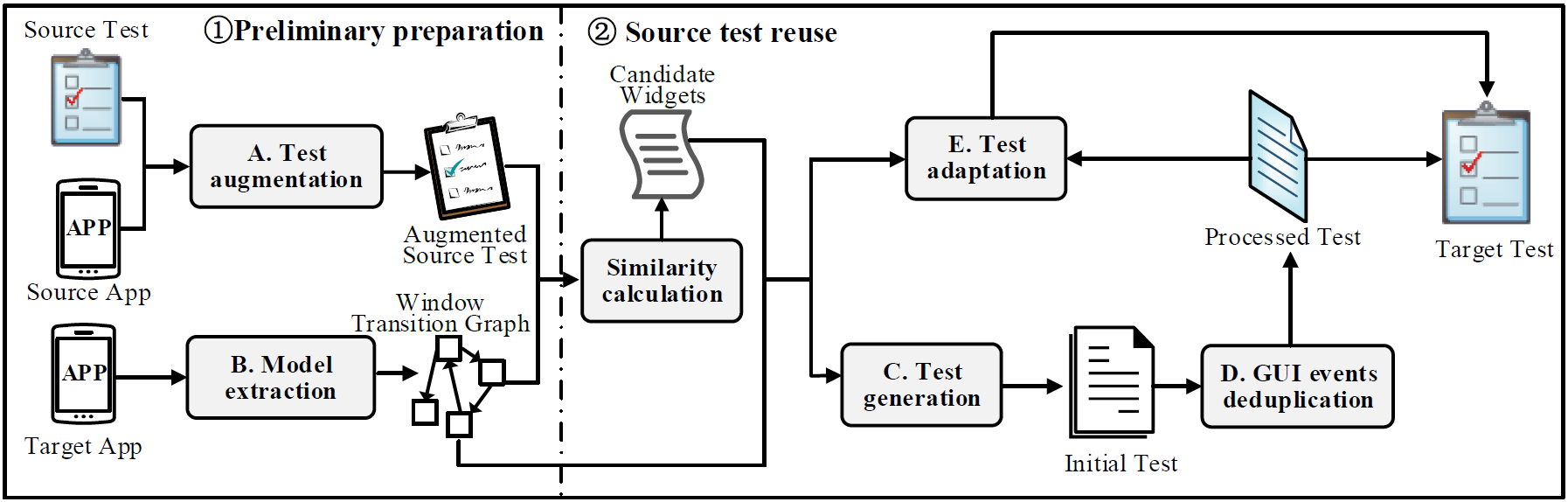}
\centering
\caption{ The overview of TRASM. \label{fig2}}
\end{figure*}

Figure~\ref{fig2} shows an overview of the proposed test reuse approach TRASM. Based on semantic matching of events, TRASM considers the  test ({\itshape source test}) of the existing application ({\itshape source app}), and the new application ({\itshape target app}) as inputs and outputs target test. TRASM employs two significant phases to implement test reuse: preliminary preparation and source test reuse. For the former,
the existing data is processed through test augmentation and model extraction to facilitate the implementation of source test reuse. For the latter, the processed data obtained by the former is used together to reuse the source test on the target app.

Test augmentation and model extraction are preliminary preparation steps that follow existing work \cite{lin2019test}, and we will only briefly introduce them. In detail, we focus on introducing our main contributions.

\subsection{Test augmentation}
The main task of test augmentation is to extract semantic information of widgets during the execution of collected source tests. The semantically represented widgets, together with actions, compose augmented tests, which are used to match widgets in the GUI of the target app.

After the source app executes each event, the \emph{adb} tool\footnote{\url{https://developer.android.com/studio/command-line/adb.html}} is used to extract the semantic information of the corresponding widget in the executed event according to the reached GUI state.  Multiple attributes (including class, resource-id, text, content-desc, clickable, password, parent\_text, sibling\_text, activity, and package) uniquely represent a widget in the GUI. These non-empty attributes and their values constitute the widget's semantic information. For example, for widget $w_{1}^{t}$ of test (a) in Figure 1, the collected semantic information is shown in Table 1.

\begin{table}[htbp]
\centerline { Table 1. The semantic information of widget $w_{1}^{t}$.}
\vskip2pt
\begin{center}
\begin{tabular}{cc}
  \hline
  \bf {Attribute} &\bf {Value} \\
   \hline
  class & android.widget.ImageButton  \\
  resource-id & fab\_new\_task \\
  clickable&true  \\
  password & false\\
  activity& .view.MainActivity  \\
  package & org.secuso.privacyfriendlytodolist\\

  \hline
\end{tabular}
\end{center}
\end{table}

\subsection{Model extraction}
The model extraction aims to statically analyze the source code obtained by the target app to obtain the window transition graph (WTG). Following with existing work, we employ  tool\footnote{\url{https://drive.google.com/file/d/1HEFS96c5nNKnzBPkWlRdwBiunOHgOs-/view?usp=sharing}} to obtain the WTG. And the steps of constructing WTG can refer to literature \cite{lin2019test}.

\begin{figure}[htbp]
\includegraphics[scale=0.25]{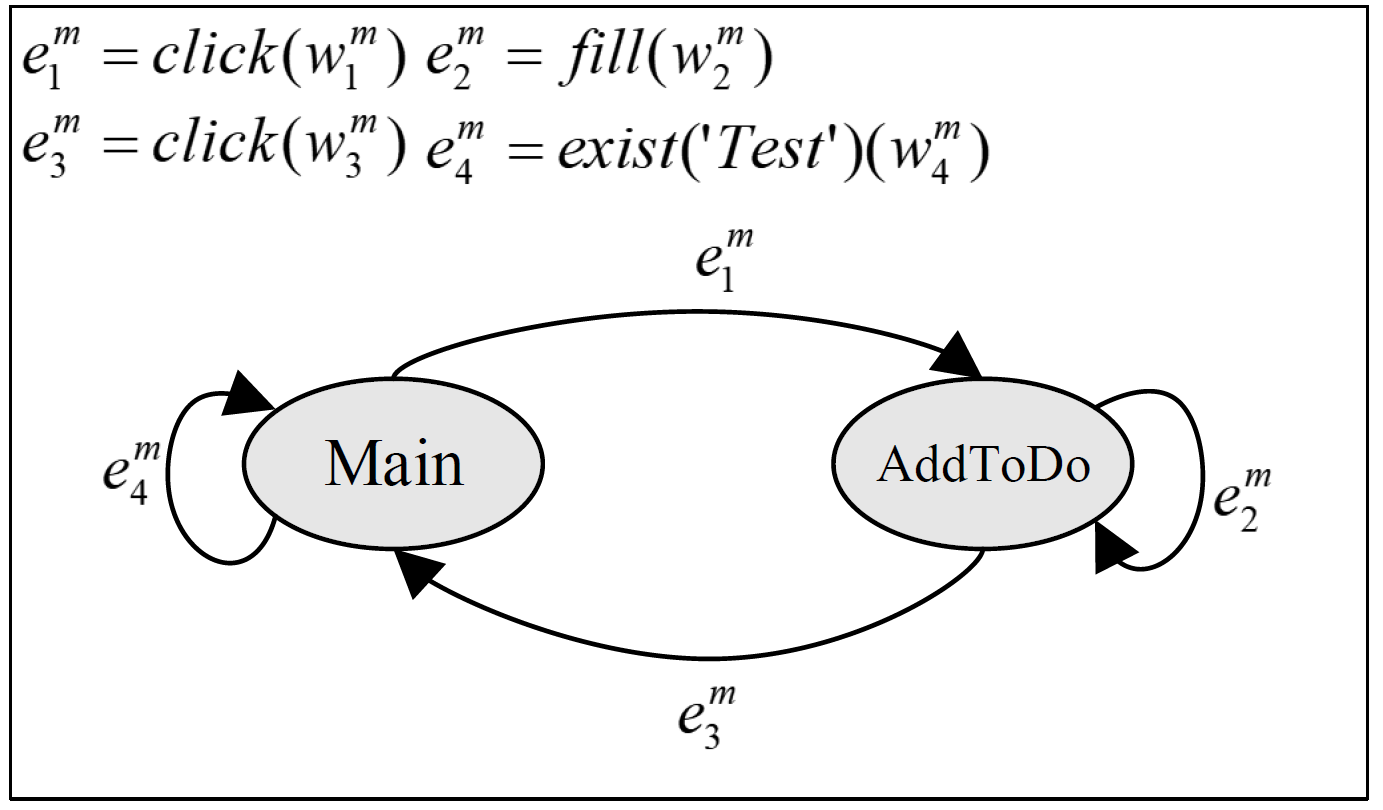}
\centering
\caption{ Window transition triggered by test (b). \label{fig3}}
\end{figure}

The WTG can visually represent the interaction between application activities and is composed of node sets and edges. Among WTG, the node represents the activity of the application, and the directed edge represents the activity transition that the event can trigger. For example, for the test (b) of application {\itshape Minimal} in Figure~\ref{fig1}, the window transition triggered by the execution of the event is shown in Figure~\ref{fig3}, where the nodes Main and AddToDo respectively represent the two activities of the application {\itshape Minimal}. By triggering the event on edge, state transition occurs between activities. The obtained WTG can provide matching candidate widgets for widgets in the source test. However, the WTG obtained may be incomplete. More fully, we adopt updating the WTG based on the feedback running information when executing the application.

\subsection{Test generation}\label{tgen}
The purpose of test generation is to generate the initial test on the target app according to the semantic information of the augmented test and the WTG of the target app.

Every event in the augmented test iteratively matches the corresponding events in the target app. The candidate widgets are first obtained by similarity calculation to match the event.\\
\textbf{Similarity calculation.} For each widget, the semantic information is captured from the current GUI page of the target app to build a word list of attributes. For example, the attribute `resource-id' of widget $w_{2}^{t}$ and widget $w_{2}^{m}$ are `$et\_new\_task\_name$' and `$userToDoEditText$' respectively. After preprocessing~\cite{lin2019test,zhou2020user}, we get two word lists
$a$=[`edit', `text', `new', `task', `name'] and $a'$=[`user', `todo', `edit', `text']. For any word $w\in a$ and $w' \in a'$, the highest similarity $\max sim(w,w')$ between words $w$ and $w'$  is synthesized as the similarity of attributes:
\begin{equation}
sim(a,a')=\frac{\sum_{w\in a}\max_{w'\in a'}sim(w,w')}{|a|}
\end{equation}

where $a$ and $a'$ represent the attributes corresponding to widget $S$ in source test and widget $T$ in target app respectively, $sim(w,w')$ expresses the cosine distance of the word vectors $\overrightarrow{V_{w}}$ and $\overrightarrow{V_{w'}}$, obtained by the Word2Vec model\cite{mikolov2013distributed}:
\begin{equation}
sim(w,w')=\frac{\overrightarrow{V_{w}}\cdot\overrightarrow{V_{w'}}}{|\overrightarrow{V_{w}}||\overrightarrow{V_{w'}}|}
\end{equation}

Based on Equation (1), we get the similarity between widget $S$ and widget $T$:

\begin{equation}
sim(S,T)=\frac{\sum_{a\in S}sim(a,a')\ast wg(a) }{|S|}
\end{equation}

where, $wg(a)$ represents the weight of attribute $a$ among all attributes. Based above calculation, we build candidate widgets by selecting several widgets in the target app with high similarity.

We identify a reachable widget based on the obtained candidate widgets and assign an action to form an event. All the paths from the current activity to the activity of each candidate widget can be queried from the WTG. These paths are executed to identify the reachable widget and return leading events. In addition, to avoid repeated path exploration, we adopt a strategy to preserve the path that has been explored and the corresponding leading events. For example, when matching event $e_{2}^{t}$ in test (a), the application {\itshape Minimal} reaches the AddToDo activity after executing event $e_{1}^{m}$ as shown in Figure ~\ref{fig3}, and then candidate widgets on the current page are collected. From the stored explored paths, it is found that there is a reachable path between activity AddToDo and activity AddToDo. Widget $w_ { 2} ^ {m}$ is located in the reachable path, which is identified as a reachable widget. Finally, according to the source event, the action is allocated to the widget $w_ { 2} ^ {m}$.

\subsection{GUI events deduplication}
Invalid repeated events will increase the complexity of test execution. Although repeated events in the test will not affect the triggering of the behavior of an application, GUI events deduplication intends to reduce the time consumption occupied by such events.

Since the GUIs of the two applications are different, the target app may not have the special functionality contained in the source test. As explained in Section ~\ref{sec1}, the reuse of special functionality in test (a), that is, the matching of event $e_{0}^{t}$ on application {\itshape Minimal}, is meaningless. To remove such GUI events in the test sequence, deduplication is performed. However, it is a challenge to identify the meaningless events in the test. We take the operation of detecting and deleting duplicate events unrelated to the generated initial test. Considering the variety of possible duplicate event patterns, we set two rules to distinguish them. First, if only a single event is repeated in the initial test, we delete the repeated events at the beginning of the test sequence. Second, if the test sequence starts with $\langle e_{n_{0}},e_{n_{1}}\rangle$ and also contains $\langle e_{n_{1}},e_{n_{0}}\rangle$ such events, we delete the pair of events. After this operation, to maintain the correctness, we check whether the test after deduplication, that is, the processed test, can maintain the functionality as the initial test. If not, we will give up the GUI events deduplication.

\subsection{Test adaptation}

The goal of test adaptation is to explore whether there is a better test sequence than the processed test using the designed adaptation strategy.

Test generation always prioritizes the widget with the highest similarity for matching. When the method of calculating similarity does not work well, it may not be possible to distinguish the best widgets to match, which will affect the accuracy of the result. The design idea of test adaptation is to find indexes that may have more semantically similar events in the processed test and then rematch them. However, determining such indexes in the sequences of the processed test is a challenging task. In this paper, we first record the indexes for which widgets in the processed test have higher similarity to another widget in the augmented test, except for the current matching event. Then, we choose the indexes with the lowest similarity of matching events in the processed test, which tries to mine the event with the incorrect match. After these two processing stages, we obtain the index sets of events that can perform rematching. Based on the above, we successively rematch the events of each index set from the candidate widgets obtained by Section~\ref{tgen}. We set the early termination condition to obtain a new test sequence that is more semantically similar than the original ones.

We explain how test adaptation solves the problem of reusing test (b) to application {\itshape To-Do List} in Figure ~\ref{fig1}. As mentioned in Section ~\ref{sec1}, different GUI designs make the similarity between the correct widget and the source widget low, resulting in the incorrect match of event $e_{3}^{m}$. Through the strategies mentioned above, we get the index of event $e_{3}^{m}$ to be rematched. Then, combined with the WTG obtained from the model extraction, the correct reachable widget $w_{3}^{t}$ is searched again from the obtained candidate widgets on this index to form event $e_{3}^{t}$. Finally, the process ends after the oracle event $e_{4}^{t}$ matching.

%% file: sections/Evaluation.tex
\section{Experimental evaluation}
We implement our approach TRASM as a tool. Moreover, we compared TRASM with the baseline approach CRAFTDROID \cite{lin2019test}, a test transfer method across mobile applications through semantic mapping, to verify the effectiveness and efficiency of TRASM. In this section, we introduce the experimental setup and experimental results to evaluate TRASM.

\subsection{Experimental setup}

For consistency, we reused the dataset\footnote{\url{https://sites.google.com/view/craftdroid/}} of \cite{lin2019test} to evaluate the proposed TRASM. Following the steps of the baseline, we conducted reuse tests on 15 applications in three categories, including {\itshape browser}, {\itshape Tip Calculator}, and {\itshape To-Do List}. These applications come from Google Play and F-Droid, which are often used in the GUI testing field to explore the functionalities of application \cite{mao2016sapienz,rau2018transferring,choudhary2015automated}. Concretely, Table 2 details the category, name (version), and source of each application.

\begin{table}[ht]
\centerline { Table 2. The specific information of applications.}
\vskip2pt
\begin{center}
\resizebox{.95\columnwidth}{!}{
\begin{tabular}{clc}
  \hline
  {\bf Category}&{\bf \hspace{1cm} Application (version)}&{\bf \hspace{0.2cm} Source} \\
   \hline
  \multirow{5}{*}{\hspace{0cm}a1-Browser} & a11-Lightning (4.5.1) & F-Droid \\
   &a12-Browser for Android (6.0) &Google Play \\
   & a13-Privacy Browser (2.10) & F-Droid \\
   & a14-FOSS Browser (5.8) & F-Droid \\
   & a15-Firefox Focus (6.0)& Google Play \\
    \hline
  \multirow{5}{*}{\hspace{0cm}a2-Tip Calculator} & a21-Tip Calculator (1.1) &Google Play \\
   & a22-Tip Calc (1.11) & Google Play\\
  & a23-Simple Tip Calculator (1.2) & Google Play \\
  & a24-Tip Calculator Plus (2.0) & Google Play \\
  &a25-Free Tip Calculator (1.0.0.9) & Google Play \\
   \hline
  \multirow{5}{*}{\hspace{0cm}a3-To Do List} & a31-Minimal (1.2) & F-Droid \\
   & a32-Clear List (1.5.6) & F-Droid \\
   & a33-To-Do List (2.1) & F-Droid \\
   & a34-Simply Do (0.9.1) & F-Droid \\
   & a35-Shopping List (0.10.1) & F-Droid\\
  \hline
\end{tabular}}
\end{center}
\end{table}

Specifically, for each application category, two typical functionalities are selected, and the corresponding tests of each application are collected according to the functionalities. To achieve the goal of verifying the implemented functionality, the last event of each test case is set as an oracle. In general, there are six functionalities in three categories of applications, as shown in Table 3. Table 3 lists the number of test cases for each functionality and the average number of GUI and oracle events.

\begin{table}[ht]
\centerline { Table 3. Tests for the typical functionalities.}
\vskip2pt
\begin{center}
\resizebox{.85\columnwidth}{!}{
\begin{tabular}{|l|c|c|c|}
  \hline
  \multirow{2}{*}{\hspace{1.2cm}\bf Functionality} &\bf Test& \bf Avg & \bf Avg \\
   &\bf Cases& \bf GUIs & \bf Oracles \\
   \hline

  b11-Access website by URL & 5 &3.4 & 1 \\
  b12-Back button & 5 & 7.4 & 3 \\
  b21-Calculate total bill with tip&5 & 3.8 & 1 \\
 b22-Split bill& 5 &4.8 & 1 \\
  b31-Add task& 5 &4 & 1 \\
  b32-Remove task &5 & 6.8 & 2 \\
  \hline
  \hspace{1.5cm}Total & 30 & 5.1 & 1.5 \\
  \hline
\end{tabular}}
\end{center}
\end{table}

Our experiment was implemented on a Nexus 5X Emulator running Android 6.0 (API 23) installed on a Ubuntu desktop with a 3.4 GHz Intel Core i7 CPU and 32 GB RAM.

\subsection{Experimental results}
 \begin{table*}[ht]
\centering {Table 4. Effectiveness and Efficiency Evaluation}
\vskip2pt
\begin{tabular}{ccccccc}
\hline
\multirow{2}{*}{Functionality}  & \multirow{2}{*}{Approach} & \multicolumn{2}{c}{GUI Event} & \multicolumn{2}{c}{Oracle Event} & \multirow{2}{*}{Successful Reuse}\\
                      &     & Precision       & Recall       & Precision       & Recall      &      \\
\hline
\multirow{2}{*}{b11} & CRAFTDROID                  &79\%   & 100\%  & 100\% & 100\% &$20/20(100\%)$        \\
                                    & TRASM                  & \textbf{100}\%  & 100\% & 100\% & 100\% &$20/20(100\%)$  \\
\hline
\multirow{2}{*}{b12} & CRAFTDROID                  &85\%   & 100\%  & 100\% & 100\% &$20/20(100\%)$        \\
                                    & TRASM                  & \textbf{100}\%  & 100\% & 100\% & 100\% &$20/20(100\%)$  \\
\hline
\multirow{2}{*}{b21} & CRAFTDROID                  &82\%   & 100\%  & 100\% & 80\% &$16/20(80\%)$        \\
                                    & TRASM                 & \textbf{93}\%  & 100\% & 100\% & \textbf{90}\% &$\textbf{18/20(90\%)}$  \\
\hline
\multirow{2}{*}{b22} & CRAFTDROID                  &80\%   & \textbf{100\%}  & 100\% & 65\% &$13/20(65\%)$        \\
                                    & TRASM                  & \textbf{85}\%  & 98\% & 100\% & \textbf{75\% }&$\textbf{15/20(75\%)}$  \\
\hline
\multirow{2}{*}{b31} & CRAFTDROID                  &78\%   & 100\%  & 85\% & 100\% &$17/20(85\%)$       \\
                                    & TRASM                  &\textbf{87}\%   &  100\%  & 85\% & 100\% &$17/20(85\%)$  \\
\hline
\multirow{2}{*}{b32} & CRAFTDROID                  &69\%   & \textbf{100\%}  & 85\% & 80\% &$11/20(55\%)$         \\
                                    & TRASM                  &\textbf{81}\%   &97\%  &\textbf{93}\%  & \textbf{81}\% &$\textbf{12/20(60\%)}$  \\
\hline
\end{tabular}
\end{table*}
This subsection presents the experimental results of TRASM and the baseline approach CRAFTDROID under the same evaluation metrics.  For each functionality of each category, we reuse the test of one application on the remaining four applications respectively, and the total number of test reuse is 5(test cases) $\times$ 4(target applications)  = 20. This paper shows the average result of 20 different test reuses. In order to avoid randomness, for each test reuse, we take the average of the multiple results recorded.

 \textbf{Effectiveness.} By comparison, the tests reused by TRASM perform higher usability than CRAFTDROID. The following two aspects, including the evaluation of successful reuse and the evaluation of matching events, can support the usability of the TRASM approach reuse test.

 Regarding reuse success rate, TRASM has significantly improved test reuse in 3 of the six functionalities, as shown in the last column of Table 4. For functionalities b21 and b22, successfully reused tests achieved a 10\% increase. For functionality b32, successful reuse also increased from 20\% to 25\%. In addition, 2 of the six functionalities, namely b11 and b12, have shown the highest successful reuse, i.e., 100\%, no matter whether it is approach CRAFTDROID or TRASM.

 For the evaluation of matching events, the third and fourth columns of Table 4 list the precision and recall of GUI events and oracle events, respectively. As shown in the table, compared with CRAFTDROID, TRASM improves the precision of the GUI by 5\% to 15\% in different functionalities. Unfortunately, while improving the precision, the recall rate of GUI events for functionalities b22 and b32 has decreased slightly by 2\% and 3\%, respectively. The success of the reused test depends on whether the match of the last oracle event in the test sequence is correct. Therefore, the improvement of successful reuse also represents the increase in the recall rate of oracle events, as listed in Table 4. Among them, the most significant is that for functionalities b21 and b22, the recall rate is improved by 10\%. In general, the improvement in the precision and the recall of oracle events shows that the proposed TRASM indeed increases the availability of the reused test.

\begin{figure}[htbp]
\includegraphics[scale=0.3]{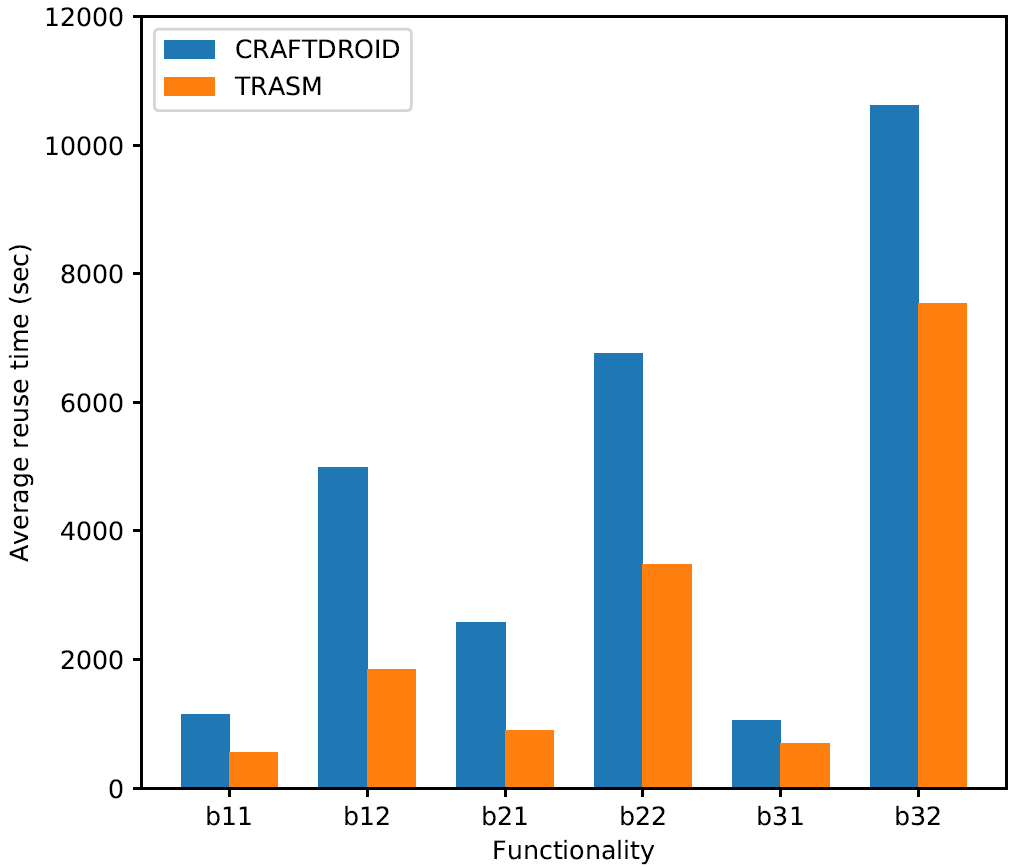}
\centering
\caption{The average test reuse time on each functionality.\label{fig4}}
\end{figure}

  \textbf{Efficiency.} Figure ~\ref{fig4} lists the average test reuse time on each functionality. It is obvious that the average time spent on reuse testing of TRASM is less than that of CRAFTDROID for each functionality. Even the most significant effect is that for functionality b21, the average reuse test of CRAFTDROID takes 2581 seconds (43 minutes), while our TRASM only takes 890 seconds (15 minutes), which is close to 35\% of the time of CRAFTDROID. In summary, the results are attributed to two factors. One is that the storage of explored paths avoids repeated time consumption, and the other is that the adaptive strategy improves the efficiency of widget matching. The above results prove that we can break through the limitation on efficiency in CRAFTDROID.

  While evaluating the efficiency of TRASM, an important finding is that implementing an adaptive strategy can improve the success of some reused tests. Inevitably, the potential drawback is that more suitable event matching can not be found will bring additional time consumption. We need to address this crucial point further to balance efficiency and effectiveness.


%% file: sections/Conclusion.tex
\section{Conclusion}
Test reuse as an alternative method of test generation can help developers verify the behavior of applications. In this paper, a novel test reuse approach has been proposed to alleviate the challenge of semantic problems in event matching.  
From the initial test set, we have extended GUI events deduplication and test adaptation to build up target tests. The experimental results indicate that our proposed approach achieves better performance than the baseline approaches with increased usability of the reused tests.

We believe that matching events 
are a promising direction, and we plan to study how to improve the matching strategy further in the future. In addition, we plan to verify the generalization of the method and further explore the effect of test reuse on more applications.

%% file: main.bbl
\begin{thebibliography}{1}


\bibitem{anand2012automated}
S. Anand, M. Naik, M. J. Harrold, and H. Yang, ``Automated concolic testing of
  smartphone apps,'' \emph{Proceedings of the ACM SIGSOFT 20th International
  Symposium on the Foundations of Software Engineering}, pp. 1--11, November 2012.

\bibitem{joorabchi2013real}
M. E. Joorabchi, A. Mesbah, and P. Kruchten, ``Real challenges in mobile app
  development,'' \emph{2013 ACM/IEEE International Symposium on Empirical
  Software Engineering and Measurement}.\hskip 1em plus 0.5em minus 0.4em\relax
  IEEE, pp. 15--24, October 2013.

\bibitem{kochhar2015understanding}
P. S. Kochhar, F. Thung, N. Nagappan, T. Zimmermann, and D. Lo, ``Understanding
  the test automation culture of app developers,'' \emph{2015 IEEE 8th
  International Conference on Software Testing, Verification and Validation}.\hskip 1em plus 0.5em minus 0.4em\relax IEEE, pp. 1--10, April 2015.

\bibitem{linares2017developers}
M. Linares-V{\'a}squez, C. Bernal-C{\'a}rdenas, K. Moran, and D. Poshyvanyk,
  ``How do developers test android applications?'' in \emph{2017 IEEE
  International Conference on Software Maintenance and Evolution
  (ICSME)}.\hskip 1em plus 0.5em minus 0.4em\relax IEEE, pp. 613--622, September 2017.

\bibitem{mao2016sapienz}
K. Mao, M. Harman, and Y. Jia, ``Sapienz: Multi-objective automated testing for
  android applications,'' \emph{Proceedings of the 25th international
  symposium on software testing and analysis}, pp. 94--105, July 2016.

\bibitem{machiry2013dynodroid}
A. Machiry, R. Tahiliani, and M. Naik, ``Dynodroid: An input generation system
  for android apps,'' \emph{Proceedings of the 2013 9th Joint Meeting on
  Foundations of Software Engineering}, pp. 224--234, August 2013.

\bibitem{gu2019practical}
T. X. Gu, C. N. Sun, X. X. Ma, C. Cao, C. Xu, Y. Yao, Q. R. Zhang, J. Lu, and Z. D. Su,
  ``Practical gui testing of android applications via model abstraction and
  refinement,'' \emph{2019 IEEE/ACM 41st International Conference on
  Software Engineering}.\hskip 1em plus 0.5em minus 0.4em\relax IEEE Computer Society, pp. 269--280, August 2019.



\bibitem{mirzaei2015sig}
N. Mirzaei, H. Bagheri, R. Mahmood, and S. Malek, ``Sig-droid: Automated system
  input generation for android applications,'' \emph{2015 IEEE 26th
  International Symposium on Software Reliability Engineering}.\hskip
  1em plus 0.5em minus 0.4em\relax IEEE Computer Society, pp. 461--471, November 2015.

\bibitem{ermuth2016monkey}
M. Ermuth and M. Pradel, ``Monkey see, monkey do: Effective generation of gui
  tests with inferred macro events,'' \emph{Proceedings of the 25th
  International Symposium on Software Testing and Analysis}, pp. 82--93, July 2016.
  
\bibitem{zhou2020user}
Y. Zhou, Y. Su, T. Chen, Z. Huang, H. C. Gall, and S. Panichella, ``User
  review-based change file localization for mobile applications,'' \emph{IEEE
  Transactions on Software Engineering}, vol. 47, no. 12, pp. 2755--2770, January 2020.

\bibitem{dong2020time}
Z. Dong, M. B{\"o}hme, L. Cojocaru, and A. Roychoudhury, ``Time-travel testing
  of android apps,'' \emph{2020 IEEE/ACM 42nd International Conference on
  Software Engineering}, Seoul, South Korea, pp. 481--492, July 2020.


\bibitem{memon2003gui}
A. Memon, I. Banerjee, and A. Nagarajan, ``Gui ripping: Reverse engineering of
  graphical user interfaces for testing,'' \emph{10th Working Conference on
  Reverse Engineering, 2003. WCRE 2003. Proceedings.}\hskip 1em plus 0.5em
  minus 0.4em\relax Citeseer, pp. 260--269, 2003.

\bibitem{su2017guided}
T. Su, G. Meng, Y. Chen, K. Wu, W. Yang, Y. Yao, G. Pu, Y. Liu, and Z. Su,
  ``Guided, stochastic model-based gui testing of android apps,''
  \emph{Proceedings of the 2017 11th Joint Meeting on Foundations of Software
  Engineering}, pp. 245--256, August 2017.

\bibitem{amalfitano2012using}
D. Amalfitano, A. R. Fasolino, P. Tramontana, S. De Carmine, and A. M. Memon,
  ``Using gui ripping for automated testing of android applications,''
  \emph{2012 Proceedings of the 27th IEEE/ACM International Conference on
  Automated Software Engineering}.\hskip 1em plus 0.5em minus 0.4em\relax IEEE, pp. 258--261,
 September 2012.


\bibitem{wang2020combodroid}
J. Wang, Y. Y. Jiang, C. Xu, C. Cao, X. X. Ma, and J. Lu, ``Combodroid: generating
  high-quality test inputs for android apps via use case combinations,''
  \emph{Proceedings of the ACM/IEEE 42nd International Conference on Software
  Engineering}, pp. 469--480, June 2020.



\bibitem{behrang2018automated}
F. Behrang and A. Orso, ``Automated test migration for mobile apps,''
  \emph{Proceedings of the 40th International Conference on Software
  Engineering: Companion Proceeedings}, pp. 384--385, May 2018.

\bibitem{behrang2019test}
F. Behrang and A. Orso, ``Test migration between mobile apps with similar functionality,''
  \emph{2019 34th IEEE/ACM International Conference on Automated Software
  Engineering}.\hskip 1em plus 0.5em minus 0.4em\relax IEEE Computer Society, pp.
  54--65, November 2019.

\bibitem{behrang2020apptestmigrator}
F. Behrang and A. Orso, ``Apptestmigrator: a tool for automated test migration for android
  apps,'' \emph{2020 IEEE/ACM 42nd International Conference on Software
  Engineering: Companion Proceedings}.\hskip 1em plus 0.5em
  minus 0.4em\relax IEEE, pp. 17--20, October 2020.

\bibitem{lin2019test}
J. W. Lin, R. Jabbarvand, and S. Malek, ``Test transfer across mobile apps
  through semantic mapping,'' \emph{2019 34th IEEE/ACM International
  Conference on Automated Software Engineering}, pp. 42--53, November 2019.

\bibitem{mao2021user}
Q. Mao, W. W. Wang, F. You, R. L. Zhao, and Z. Li, ``User behavior pattern mining and
  reuse across similar android apps,'' \emph{Journal of Systems and Software},
  vol. 183, pp. 111085,  September 2021.

  \bibitem{mariani2021semantic}
L. Mariani, A. Mohebbi, M. Pezz{\`e}, and V. Terragni, ``Semantic matching of
  gui events for test reuse: are we there yet?'' \emph{Proceedings of the
  30th ACM SIGSOFT International Symposium on Software Testing and Analysis}, pp. 177--190,
 July 2021.

\bibitem{qin2019testmig}
X. Qin, H. Zhong, and X. Y. Wang, ``Testmig: Migrating gui test cases from ios to
  android,'' \emph{Proceedings of the 28th ACM SIGSOFT International
  Symposium on Software Testing and Analysis}, pp. 284--295, July 2019.

\bibitem{rau2018poster}
A. Rau, J. Hotzkow, and A. Zeller, ``Poster: Efficient gui test generation by
  learning from tests of other apps,'' \emph{2018 IEEE/ACM 40th
  International Conference on Software Engineering: Companion
  (ICSE-Companion)}.\hskip 1em plus 0.5em minus 0.4em\relax IEEE, pp.
  370--371, June 2018.

\bibitem{rau2018transferring}
A. Rau, J. Hotzkow, and A. Zeller, ``Transferring tests across web applications,'' \emph{International
  Conference on Web Engineering}.\hskip 1em plus 0.5em minus 0.4em\relax
  Springer, pp. 50--64, May 2018.

\bibitem{choudhary2015automated}
S. R. Choudhary, A. Gorla, and A. Orso, ``Automated test input generation for
  android: Are we there yet?(e),'' \emph{2015 30th IEEE/ACM International
  Conference on Automated Software Engineering}.\hskip 1em plus 0.5em
  minus 0.4em\relax IEEE, pp. 429--440, November 2015.

\bibitem{zeng2016automated}
X. Zeng, D. Li, W. Zheng, F. Xia, Y. Deng, W. Lam, W. Yang, and T. Xie,
  ``Automated test input generation for android: Are we really there yet in an
  industrial case?'' in \emph{Proceedings of the 2016 24th ACM SIGSOFT
  International Symposium on Foundations of Software Engineering}, pp.
  987--992, November 2016.

\bibitem{mikolov2013distributed}
T.~Mikolov, I.~Sutskever, K.~Chen, G.~S. Corrado, and J.~Dean, ``Distributed
  representations of words and phrases and their compositionality,''
  \emph{Advances in neural information processing systems}, vol.~26, 2013.




\end{thebibliography}
